\documentclass[showpacs,twocolumn,superscriptaddress,pra,aps,longbibliography]{revtex4-1}

\usepackage{braket}
\usepackage{amsmath}
\usepackage{amsthm}
\usepackage{amssymb}
\usepackage{mathrsfs}
\usepackage{mathtools}
\usepackage{color}
\usepackage{comment}

\makeatletter

\newcommand{\Rmnum}[1]{\expandafter\@slowromancap\romannumeral #1@}
\makeatother

\DeclarePairedDelimiter\floor{\lfloor}{\rfloor}

\begin{document}

\title{Conditions for anti-Zeno-effect observation in free-space atomic radiative decay}%

\newcommand{\addrMRS}{Aix Marseille Univ, CNRS, Centrale Marseille, Institut Fresnel, Marseille, France}
\newcommand{\addrMPIK}{Max Planck Institute for Nuclear Physics, Saupfercheckweg 1, 69117 Heidelberg, Germany}

\author{Emmanuel Lassalle}
\affiliation{\addrMRS}

\author{Caroline Champenois}
\affiliation{Aix Marseille Univ, CNRS, PIIM, Marseille, France}

\author{Brian Stout}
\affiliation{\addrMRS}

\author{Vincent Debierre}
\email[]{vincent.debierre@mpi-hd.mpg.de}
\affiliation{\addrMPIK}

\author{Thomas Durt}
\affiliation{\addrMRS}

\date{\today}%

\begin{abstract}
Frequent measurements can modify the decay of an
unstable quantum state with respect to the free dynamics
given by Fermi's golden rule. In a landmark article [A. G. Kofman and
G. Kurizki, Nature (London) $\mathbf{405}$, 546 (2000)], Kofman and Kurizki concluded that in
quantum decay processes,
acceleration of the decay by frequent measurements, called the quantum anti-Zeno effect (AZE),
appears to be ubiquitous, while its counterpart, the quantum Zeno effect, is unattainable. However, up to now there have been no
experimental observations of the AZE for atomic radiative decay
(spontaneous emission) in
free space. In this work, making use of analytical results
available for hydrogen-like atoms, we find that in free space, only non-electric-dipolar
transitions should present an observable AZE, revealing that this effect is
consequently much less ubiquitous than first predicted.
We then propose an experimental scheme for AZE observation, involving
the electric quadrupole transition between $D_{5/2}$ and $S_{1/2}$
in the alkali-earth ions Ca$^+$ and Sr$^+$. The
proposed protocol is based on the stimulated Raman adiabatic passage technique which acts like a dephasing quasi-measurement. 
\end{abstract}

\maketitle

\section{Introduction} \label{sec:Intro}

One of the more peculiar features of quantum mechanics is that the
measurement process can modify the evolution of a quantum
system. The archetypes of this phenomenon are the quantum Zeno effect (QZE)
and the quantum anti-Zeno effect (AZE) \cite{kofman2000acceleration,
  facchi2001quantum}. The QZE refers to the inhibition of the decay of
an unstable quantum system due to frequent measurements \cite{MisraSudarshan}, and was observed
experimentally for the first time with trapped ions
\cite{cook1988quantum, itano1990quantum} and more recently in cold
neutral atoms \cite{patil2015measurement}. The
opposite effect, where the decay is \emph{accelerated} by frequent
measurements, was first called the AZE in
Ref.~\cite{kaulakys1997quantum}, and was discovered theoretically for
spontaneous emission in cavities \cite{kofman1996quantum,lewenstein2000quantum}, and first observed in a tunneling experiment with cold
atoms (along with the QZE) \cite{ZenoColdUnstable}, and recently with a
single superconducting qubit coupled to a waveguide cavity
\cite{harrington2017quantum}.
However, despite predictions that
the AZE should be much more ubiquitous than the QZE in radiative
decay processes \cite{kofman2000acceleration}, it has never been
observed to our knowledge for atomic radiative decay (spontaneous emission) in free
space.

Here, we investigate the case of hydrogen-like atoms, for which the exact expression of the coupling between the atom
and the free radiative field (\emph{cf.} \cite{Moses1973, Seke1994}) allows us to derive an analytical expression for the measurement-modified
decay rate. From this, we find that only non-electric-dipole transitions can
exhibit the AZE in free space (\emph{i.e.} non-dipole electric transitions \emph{and} magnetic transitions of \emph{any} multipolar
order), which drastically limits the experimental possibilities to observe
this effect. We
start with a brief review of the general formal results
about the measurement-modified decay rate in Sec.~\ref{sec:Emission},
and we then apply, in Sec.~\ref{sec:HLike}, this general framework to the case of
electronic transitions in hydrogen-like atoms to derive an analytical expression of the
measurement-modified decay rate in free space. Then, we discuss the experimental realizability of
the described phenomenon in Sec.~\ref{sec:XP}, and we identify a
potential candidate: the electric quadrupole transition between
$D_{5/2}$ and $S_{1/2}$ in Ca$^+$ or Sr$^+$.
Conclusions are finally given in Sec.~\ref{sec:Ccl}.

\section{Monitored spontaneous emission: general analysis} \label{sec:Emission}

We consider a two-level atom in free space, consisting of a ground state
$\ket{\mathrm{g}}$ and an excited state $\ket{\mathrm{e}}$ separated
by the Bohr energy
$\hbar\omega_0$, and initially prepared in $\ket{\mathrm{e}}$. Due to the coupling with the modes
of the electromagnetic (EM) reservoir,
the atom will naturally decay to the ground state
$\ket{\mathrm{g}}$, with a survival probability to stay in the
excited state $\ket{\mathrm{e}}$ given by $P(t) =
\text{exp}(-\Gamma t)$ (Wigner-Weisskopf decay \cite{cohen1989photons}). For the free dynamics
(\emph{i.e.} without measurements), the decay rate $\Gamma$ is given by the Fermi's golden rule
(FGR) \cite{cohen1989photons}, and will be denoted by $\Gamma_0$ in the following.

In Ref.~\cite{kofman2000acceleration}, Kofman and Kurizki showed 
that frequent measurements on an excited two-level atom, \emph{i.e.} repeated instantaneous projections onto the state $\ket{\mathrm{e}}$,
lead to a broadening of its energy level, analogous to collisional broadening. 
Therefore, the atom probes a larger range of EM modes in the reservoir
spectrum, and these new decay channels might
modify the dynamics. Specifically, it was shown, within the rotating-wave approximation (RWA), that if frequent
measurements are performed at short intervals $\tau$, the dynamics still follows an exponential decay,
but with a measurement-modified decay rate given by \cite{kofman2000acceleration}
\begin{equation}
\Gamma= 2\pi\int_0^{\infty}
\mathrm{d} \omega \, F_\tau\left(\omega-\omega_0\right)
R\left(\omega\right)\; .
\label{eq:art_decay}
\end{equation}
The effects of the
RWA on the QZE and AZE have been discussed in Refs.~\cite{zheng2008quantum, ai2010quantum},
showing no essential differences between the predictions made with and
without the RWA in the case of the reservoir that we shall consider
here. Moreover, for a discussion about a non-exponential decay, see Ref.~\cite{zhou2017quantum}.

\begin{figure*}[tb]
    \centering
    \includegraphics[scale=0.6]{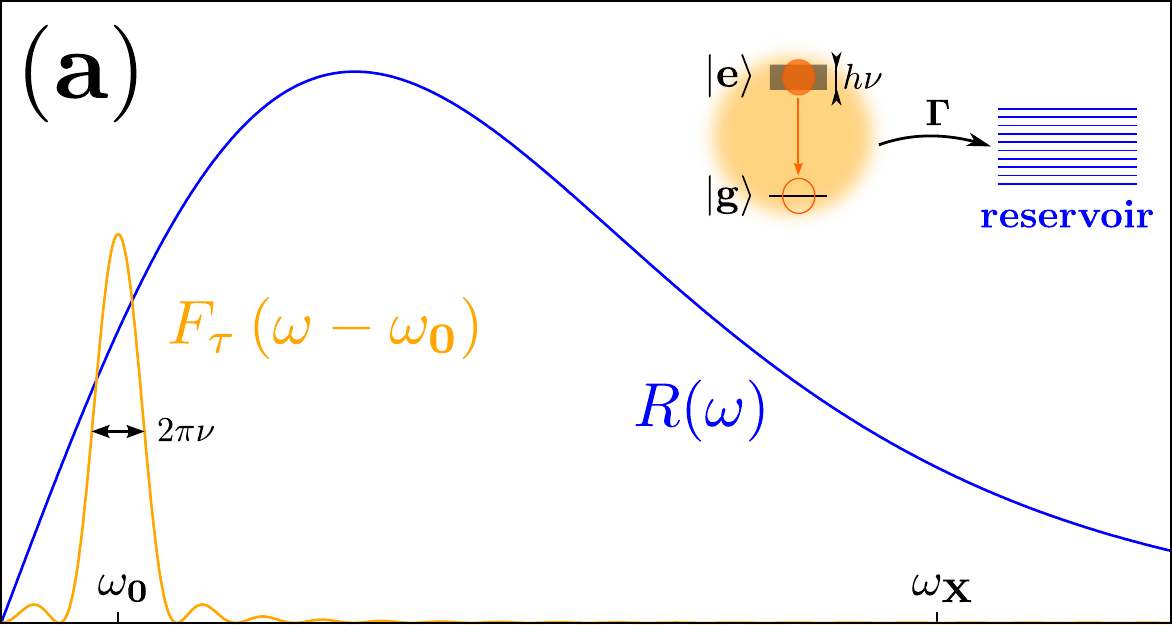}
    \hskip10mm
    \includegraphics[scale=0.6]{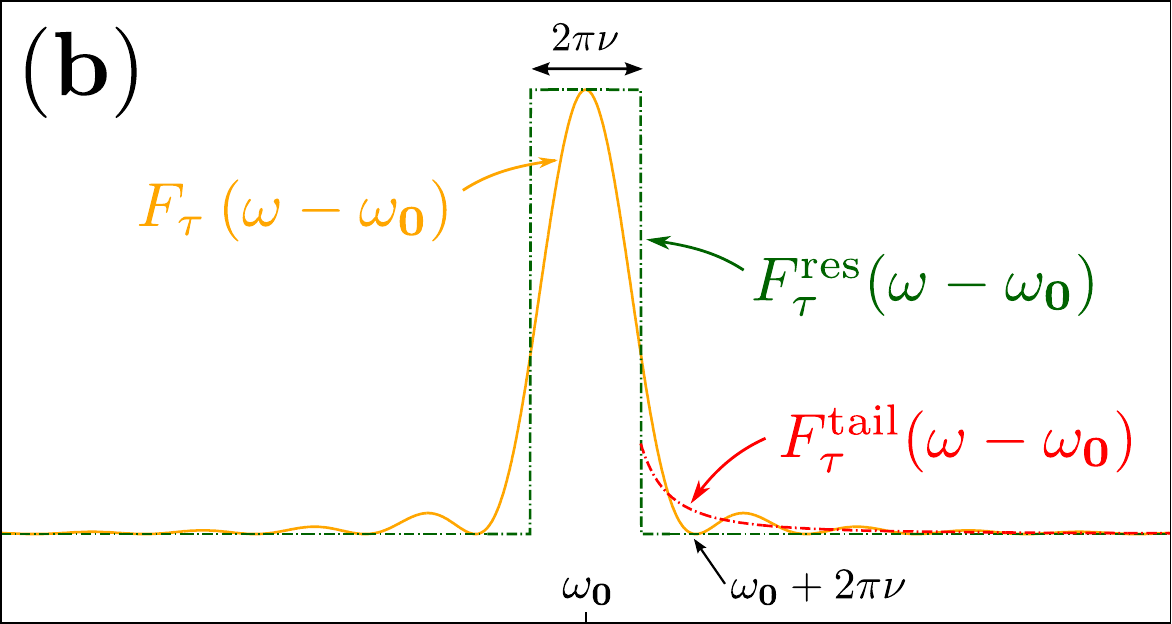}
    \caption{(a) Scheme of the broadened spectral profile
      $F_\tau(\omega-\omega_0)$ (orange line) of an atom with transition
  frequency $\omega_0$ due to repeated measurements at a rate $\nu=1/\tau$ with $\tau$ the interval between
  each measurement, and reservoir coupling spectrum
  $R(\omega)$ (blue line) of the form of Eq.~(\ref{eq:SimpleReservoir}) with a cutoff
  frequency $\omega_{\mathrm{X}}\gg\omega_0$. (b) Scheme of the broadened
  spectral profile $F_\tau(\omega-\omega_0)$ (orange line) and
  its resonant [$F^{\text{res}}_\tau(\omega-\omega_0) = 1/(2\pi\nu)$ for $-\pi\nu < \omega- \omega_0 <
\pi\nu$ (green dashed
  line)] and tail [$F^{\text{tail}}_\tau(\omega-\omega_0) =
  \nu/[\pi(\omega-\omega_0)^2]$ for $\omega- \omega_0 > \pi\nu$ (red dashed
    line)] approximations. The inset shows that the energy broadening of $\ket{\mathrm{e}}$, induced
by the frequent measurements at rate $\nu$, modifies the decay into the EM
reservoir.}
\label{fig:artistic}
\end{figure*}

In Eq.~(\ref{eq:art_decay}), the function $R(\omega)$ represents the reservoir coupling
  spectrum and is written
\begin{equation}\label{eq:reservoir}
  R(\omega) = \hbar^{-2}\sum_k |\bra{\mathrm{e},0}\hat{H}_I\ket{\mathrm{g},1_{k}}|^2\delta(\omega-\omega_k)
\end{equation}
where $\ket{\mathrm{g},1_{k}}=\ket{\mathrm{g}}\otimes\ket{1_{k}}$ is the outer product
between the atomic state $\ket{\mathrm{g}}$ and the state of the EM field
$\ket{1_{k}}$ containing one photon in the mode labelled by
$k$, $\ket{\mathrm{e},0}=\ket{\mathrm{e}}\otimes\ket{0}$ is the outer product
between the atomic state $\ket{\mathrm{e}}$ and the vacuum state of
the EM field $\ket{0}$, and $\hat{H}_I$ is the interaction Hamiltonian.
The function $F_\tau(\omega-\omega_0)$, on the other hand, corresponds to the broadened spectral profile
of the atom due to the frequent measurements at a rate $\nu=1/\tau$,
and takes the form
\begin{equation}
F_\tau\left(\omega-\omega_0\right) = \frac{\tau}{2\pi}\text{sinc}^2\left((\omega-\omega_0)\frac{\tau}{2} \right)
\label{eq:profile}
\end{equation}
with $\text{sinc}(x)\equiv\text{sin}(x)/x$. Note that the
spectral profile function can be generalized to the case where no assumption is made beforehand about
the state that is being repeatedly prepared \cite{Chaudhry}. In
Fig.~\ref{fig:artistic} (a) (orange line), the function
$F_\tau\left(\omega-\omega_0\right)$ is shown, centered on $\omega_0$
and with a width
of about $2\pi\nu$.
When $\nu \rightarrow 0$,
$F_\tau\left(\omega-\omega_0\right) \rightarrow \delta\left(\omega - \omega_0\right)$
and Eq.~(\ref{eq:art_decay}) gives: $\Gamma \rightarrow 2\pi
R(\omega_0)$, which is the natural decay
rate given by the FGR $\Gamma_0 \equiv 2\pi
R(\omega_0)$, where only the single photon states of frequency $\omega_0$
contribute to the decay.

From Eq.~(\ref{eq:art_decay}), we can see that the measurement-modified decay rate corresponds to the overlap
between the functions $R(\omega)$ and $F_\tau\left(\omega-\omega_0\right)$, and therefore depending on the profile of $R(\omega)$
in the interval around $\omega_0$, the system may experience an acceleration ($\Gamma > \Gamma_0$, AZE)
or a deceleration ($\Gamma < \Gamma_0$, QZE)  of the decay compared to the
measurement-free decay. 
In the following, we aim at investigating the case of hydrogen-like atoms
coupled to the free space EM field, for which the function $R(\omega)$ can be
calculated analytically. This will allow us to highlight the
conditions for an AZE observation in such systems. 
Before doing so, however, it is worth mentionning that in the perturbative treatment that we use, Eqs.~(\ref{eq:art_decay})
  and (\ref{eq:reservoir}) are valid to the first order (\emph{i.e.} only one-photon processes
  are considered), and do not include higher-order contributions
  (\emph{i.e.} two-photon and many-photon processes). For this approximation
  to be valid, we need to ensure that, compared to the spontaneous
  single-photon emission of the $\ket{\mathrm{e}}\rightarrow\ket{\mathrm{g}}$
  transition considered, two-photon processes, which involve
  other atomic levels, are negligible. This can only be checked on a
  case-by-case basis for specific atoms. In Sec.~\ref{sec:XP}, we consider the
  specific case of the electric quadrupole transition of Ca$^+$, and
  we check that the single-photon emission is the dominant decay channel
  from the relevant excited state (in Sec.~\ref{subsec:TransCh}).

\section{Quantum Anti-Zeno effect in hydrogen-like atoms} \label{sec:HLike}

\subsection{Reservoir coupling spectrum for hydrogen-like atoms} \label{subsec:ReservoirH}

For hydrogen-like atoms, it is useful to write the states of the
  atom in terms of the multipolar modes 
$\ket{\mathrm{g}} = \ket{n_{\mathrm{g}},l_{\mathrm{g}},m_{\mathrm{g}}}$ and $\ket{\mathrm{e}} = \ket{n_{\mathrm{e}},l_{\mathrm{e}},m_{\mathrm{e}}}$
where each atomic state is described by three discrete quantum numbers
$n_{\mathrm{i}}$, $l_{\mathrm{i}}$ and $m_{\mathrm{i}}$ which are
respectively the 
principal, angular momentum and magnetic quantum
numbers. Similarly, it is useful to write the one-photon states
in the energy-angular-momentum basis \cite{Moses1973,Seke1994} $\ket{1_{k}} = \ket{J,M,\lambda,\omega}$,
where a photon is characterized by its angular momentum and
magnetic quantum numbers $J$ and $M$, respectively, and also its helicity $\lambda$ and frequency
$\omega$. 
Based on the exact calculations of the matrix elements in (non-relativistic)
hydrogen-like atoms in \emph{free space} (initiated by Moses
\cite{Moses1973} and completed by Seke \cite{Seke1994}), the reservoir
(\ref{eq:reservoir}) can be obtained analytically and depends on the
type of the multipole transition $\ket{\mathrm{e}}\rightarrow\ket{\mathrm{g}}$
considered (see Appendix~~\ref{app:reservoir} for details)
\begin{equation} \label{eq:FullReservoir}
R(\omega) = \sum_{J=|l_\mathrm{e}-l_\mathrm{g}|}^{|l_\mathrm{e}+l_\mathrm{g}|}\sum_{r=0}^{N_J} \frac{D_{Jr}}{\omega_{\mathrm{X}}^{\eta_J+2r-1}}\frac{\omega^{\eta_J+2r}}{\left[ 1+\left(\frac{\omega}{\omega_{\mathrm{X}}}\right)^2\right]^\mu}
\end{equation}
where $\eta_J=1+2J$ for magnetic transitions, and $\eta_J=-1+2J$
for electric transitions with $J$ starts at 1 for a dipole transition ($l_\mathrm{e}-l_\mathrm{g}=1$),
at 2 for a quadrupole transition ($l_\mathrm{e}-l_\mathrm{g}=2$) and so on; $\mu =
2\left(n_{\mathrm{g}} + n_{\mathrm{e}} - 1\right)$; $D_{Jr}$ are dimensionless constants
involving the Clebsch-Gordan coefficients of the transition under consideration; and $\omega_{\mathrm{X}}$ is the
non-relativistic cutoff frequency that emerges naturally from
calculations \cite{facchi1998temporal,debierre2015spontaneous} and reads \cite{Seke1994}:
\begin{equation}\label{eq:omegacutoff}
\omega_{\mathrm{X}}=\left(\frac{1}{n_{\mathrm{g}}}+\frac{1}{n_{\mathrm{e}}}\right)\frac{c}{a_0}Z
\end{equation}
with $a_0$ the Bohr radius and $Z$ the atomic number.
Finally, the index at which the sum is
terminated is
$N_J=2\left(n_{\mathrm{e}}+n_{\mathrm{g}}\right)-4-J-l_{\mathrm{e}}-l_{\mathrm{g}}-\epsilon$
with $\epsilon=0$ for electric
transitions and $\epsilon=1$
for magnetic transitions.

For simplicity, we first consider
electric transitions ($\epsilon=0$)
between an excited state of maximal angular momentum
($l_\mathrm{e}=n_\mathrm{e}-1$) and the ground state $1S$
($n_\mathrm{g}=1$, $l_\mathrm{g}=0$). In that case, $N_J=0$ and the two
sums disappear in
Eq.~(\ref{eq:FullReservoir}) which reduces to
\begin{equation} \label{eq:SimpleReservoir}
R(\omega) = \frac{D}{\omega_{\mathrm{X}}^{\eta-1}}\frac{\omega^{\eta}}{\left[ 1+\left(\frac{\omega}{\omega_{\mathrm{X}}}\right)^2\right]^\mu}
\end{equation}
where we defined $D\equiv D_{J0}$ and $\eta\equiv \eta_J$. This
reservoir coupling spectrum is sketched on Fig.~\ref{fig:artistic} (a). The parameters $\eta$, $\mu$ and
$\omega_{\mathrm{X}}$ corresponding
to the electric transitions $2P$-$1S$ (dipole), $3D$-$1S$
(quadrupole) or $4F$-$1S$ (octupole) are given in Table~\ref{tab:coeff}.

\begin{center}
\begin{table} [t!]
\begin{center}
\def\arraystretch{1.25}
\begin{tabular}{cccc}
\hline
\hline

\multicolumn{1}{c}{Transitions} &
\multicolumn{1}{c}{$2P$-$1S$} &
\multicolumn{1}{c}{$3D$-$1S$} &
\multicolumn{1}{c}{$4F$-$1S$} \\
\hline
$\eta$ & $1$ & $3$ & $5$ \\
$\mu$ & $4$ & $6$ & $8$\\
$\omega_{\mathrm{X}}/\omega_0$ & $ 548.1 $ & $411.1 $ & $365.4 $\\
\hline
\hline
\end{tabular}
\end{center}
\caption{Parameters of the reservoir spectrum given by Eq.~(\ref{eq:SimpleReservoir}) for the electric transitions
  $2P$-$1S$ (dipole), $3D$-$1S$ (quadrupole)
or $4F$-$1S$ (octupole) in the hydrogen atom.}
\label{tab:coeff}
\end{table}
\end{center}



\subsection{Analytical results for $\omega_0\ll\omega_{\mathrm{X}}$}

In this section, we want to derive an analytical expression of the
decay rate (\ref{eq:art_decay}) to see how it scales with the measurement
rate $\nu$ when the reservoir coupling spectrum is of the form of Eq.~(\ref{eq:SimpleReservoir}), 
in the case $\omega_0\ll\omega_{\mathrm{X}}$ which is always respected for low-$Z$ atoms.
Indeed, using the Bohr formula for $\omega_0$, the ratio between $\omega_0$ and the
cutoff frequency $\omega_{\mathrm{X}}$ can be written from Eq.~(\ref{eq:omegacutoff}) as
\begin{equation} \label{eq:FreqRatio}
  \frac{\omega_0}{\omega_{\mathrm{X}}}=\frac{1}{2}\left(Z\alpha\right)\left(\frac{1}{n_{\mathrm{g}}}-\frac{1}{n_{\mathrm{e}}}\right),
\end{equation}
with $\alpha$ the fine structure constant of electrodynamics of
approximate value $\alpha \simeq 1/137$, whence we can see that the
assumption $\omega_0\ll\omega_{\mathrm{X}}$ makes sense for atoms with $Z$ moderately small. 

The details of our derivation are given in
Appendix~~\ref{app:Simple}, and we present the main ideas here.
In the integral $\Gamma$ (Eq.~(\ref{eq:art_decay})), we start by expanding (to all orders) the numerator
$\omega^\eta$ of the reservoir function $R(\omega)$ (Eq.~(\ref{eq:SimpleReservoir}))
around the transition frequency $\omega_0$. 
This binomial expansion yields a series of terms of the type $\left(\omega
-\omega_0\right)^k$ with $k$ integers between $0$ and $\eta$.
We can then consider that the total decay
rate in Eq.~(\ref{eq:art_decay}) results from two
contributions.
(i) A `resonant' contribution $\Gamma^{\text{res}}$ coming from the
$k=0$ and $k=1$ terms of the binomial expansion of $\omega^\eta$, for which only the part
of $F_\tau\left(\omega-\omega_0\right)$ that probes the reservoir $R\left(\omega\right)$
in a frequency range of width $\sim\nu$ around $\omega_0$ contributes.
This amounts to making the approximation that $F_\tau\left(\omega-\omega_0\right)=1/\left(2\pi\nu\right)$ in the interval $-\pi\nu<\omega-\omega_0<\pi\nu$ and vanishes elsewhere.
With the hierarchy $\omega_0 \ll \omega_{\mathrm{X}}$ in mind, the resonant
contribution can then be calculated
(see~Appendix~ \ref{app:Simple})
\begin{equation} \label{eq:AnotherWay}
\Gamma^{\text{res}}\simeq 2\pi\,\frac{D}{\omega_{\mathrm{X}}^{\eta-1}}\omega_0^\eta
\end{equation} 
and is found to be equal to the natural decay rate $\Gamma_0 = 2\pi R(\omega_0)
\simeq 2\pi D \omega_0^\eta/\omega_{\mathrm{X}}^{\eta-1}$ computed by
the FGR. (ii) The `tail' contribution $\Gamma^{\text{tail}}$, which only exists
if $\eta>1$, comes from all the
terms with order $1<k\leq\eta$,
for which $F^{\text{tail}}_{\tau}(\omega-\omega_0)\propto
1/(\omega - \omega_0)^2$ probes the entire reservoir and has a
non-negligible contribution. By approximating the square sine by
its mean value $1/2$, we can then
compute the tail contribution (see~Appendix~ \ref{app:Simple})
\begin{equation} \label{eq:GivesBeta}
\Gamma^{\text{tail}}\simeq  D\,\nu\,B\left(\frac{1-\eta}{2}+\mu,-\frac{1-\eta}{2}\right)
\end{equation}
where $B$ refers to Euler's Beta function and is a simple numerical
prefactor (roughly of the order of unity).
Finally, the measurement-modified decay rate $\Gamma =
\Gamma^{\text{res}}+\Gamma^{\text{tail}}$ normalized by the natural decay rate $\Gamma_0$ yields
the result (partially obtained in Ref.~\cite{kofman2000acceleration}):
\begin{multline} \label{eq:JoueLaCommeKKmain}
  \frac{\Gamma}{\Gamma_0}\simeq\left\{
    \begin{array}{ll}
       1 \text{ for } \eta=1,\\[2mm]
       1+\frac{1}{2\pi}\frac{\nu}{\omega_0}
  \left(\frac{\omega_{\mathrm{X}}}{\omega_0}\right)^{\eta-1}B\left(\frac{1-\eta}{2}+\mu,-\frac{1-\eta}{2}\right)\\
       \quad\quad\quad\quad\quad\quad\quad\quad\quad\quad\text{ for } \eta>1.
    \end{array}
\right.
\end{multline}
In Appendix~ \ref{app:Complete}, we show how this expression
can be extended to the general form of $R(\omega)$ given by
Eq.~(\ref{eq:FullReservoir}): the result is similar in terms of
scaling with the different parameters $\eta$, $\nu$, $\omega_0$
and $\omega_{\mathrm{X}}$; and the Beta function is
simply replaced by a more complex numerical prefactor (see Eq.~(\ref{eq:FinalRatio})).

\subsection{Comparison between numerical and analytical calculations
  and discussion}

\begin{figure}[tb]
  \centering
  \includegraphics[scale=0.6]{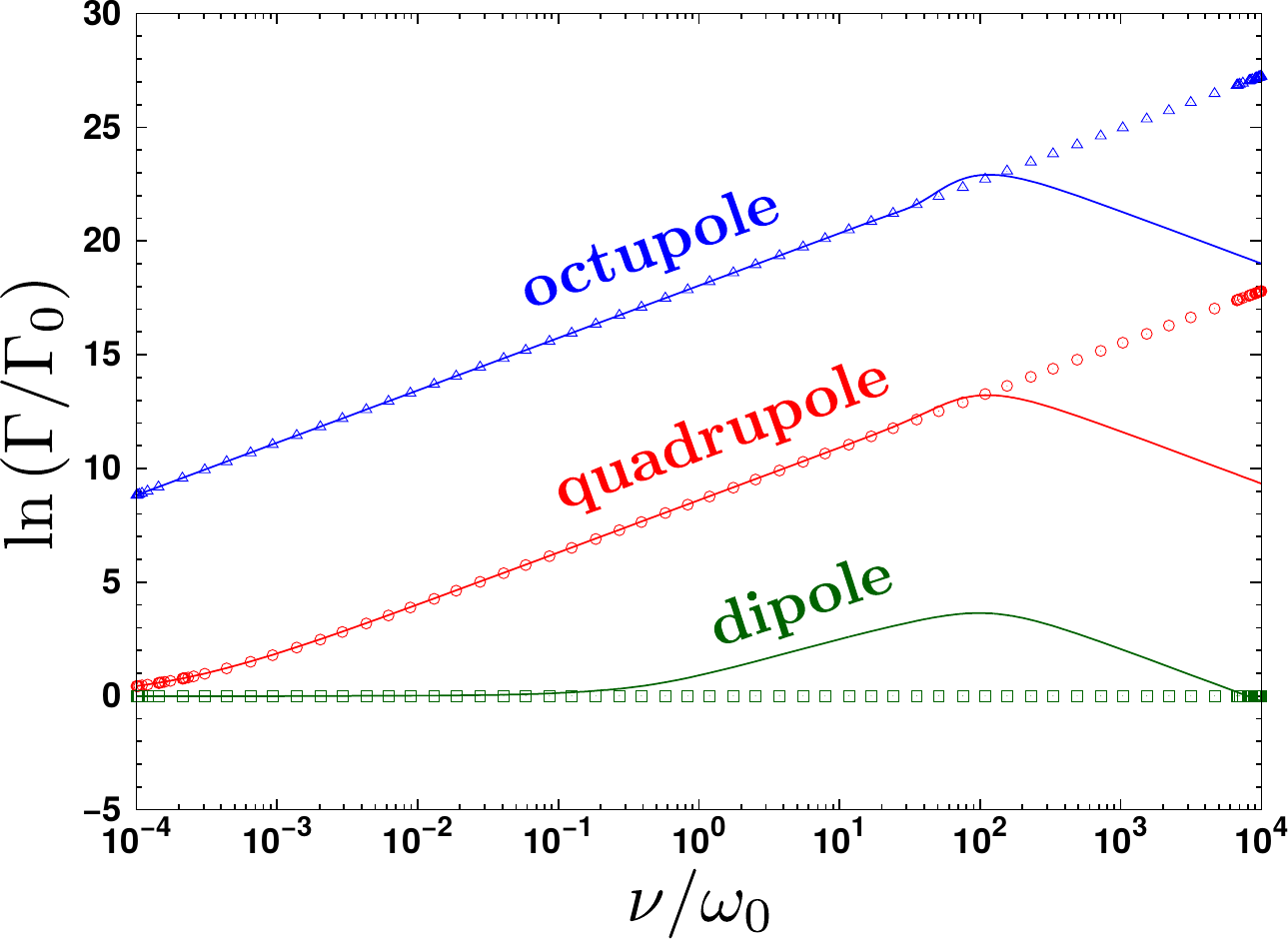}
  \caption{Comparison between numerical (full
    lines) and analytical (dotted lines) calculations of $\ln(\Gamma/\Gamma_0)$ as a
  function of the normalized measurement rate $\nu/\omega_0$ for three
  different electric transitions: dipole ($\eta=1$, in green),
  quadrupole ($\eta=3$, in red) and octupole ($\eta=5$, in blue).  
The associated
  parameters used for the function $R(\omega)$ corresponding to these
  transitions are displayed in Table~\ref{tab:coeff}.}
  \label{fig:comparison}
\end{figure}

Before commenting on the scope of this result, we first
compare in Fig.~\ref{fig:comparison} the
analytical approximation of $\Gamma$ given by
Eq.~(\ref{eq:JoueLaCommeKKmain}) to the numerical computation
$\Gamma^{\text{num}}$ of
Eq.~(\ref{eq:art_decay}) (using (\ref{eq:profile}) and
(\ref{eq:SimpleReservoir})) for three different
reservoir coupling spectra $R(\omega)$ corresponding to the electric
dipole ($\eta=1$, in green), quadrupole ($\eta=3$, in red) and octupole
($\eta=5$, in blue) transitions whose parameters are given
in Table~\ref{tab:coeff}. We can see a very
good agreement for the quadrupole and octupole transitions
up to $\nu\lesssim 100\,\omega_0$, and for the
dipolar transition up to $\nu\lesssim \omega_0$. Note that in practice, it may not be feasible to reach such high
measurement rates as $\nu\sim\omega_0$
(particularly for optical transitions, \emph{cf.} Sec.~\ref{sec:XP}),
and moreover, for $\nu \gtrsim
\omega_0$, the RWA is not valid anymore. Therefore, the analytical results
are revealed to be excellent in the regime of interest $\nu\ll\omega_0$ with
a relative error $(\Gamma^{\text{num}} -
\Gamma)/\Gamma^{\text{num}}$ less than $2\%$ for
$\nu/\omega_0<10^{-2}$ in the
three cases represented on the plot. 

Concerning the AZE, we can see that in the case of the electric dipole transition, the
AZE trend ($\Gamma>\Gamma_0$) appears only for $\nu \gtrsim
\omega_0$ (green curve) --- which is not interesting for experimental
observations as just
discussed, whereas for the other transitions (red and blue
curves), the AZE is obtained already for $\nu \ll
\omega_0$ and can be very strong. This has been overlooked in the past
and constitutes our main result:
within the natural hierarchy $\omega_0 \ll \omega_{\mathrm{X}}$, we predict from
our general Eq.~(\ref{eq:JoueLaCommeKKmain}) that electric dipole
transitions ($\eta=1$) will \emph{not} exhibit the
AZE, whereas the AZE can be expected for all other types of electronic
transitions ($\eta>1$). On the one hand, as electric dipole transitions
are arguably the most standard and studied type of electronic
transitions in atoms, these predictions make the AZE much less ubiquitous
than what had been stated in Ref.~\cite{kofman2000acceleration}.
On the other hand, we see from Eq.~(\ref{eq:JoueLaCommeKKmain}) that
for all other transitions, the ratio $\omega_{\mathrm{X}}/\omega_0\gg1$
may give rise, despite the ratio $\nu/\omega_0\ll1$, to a
strong anti-Zeno effect $\Gamma\gg\Gamma_0$, particularly for
high-order multipolar transitions. The goal of the next section is to identify realistic systems suitable
for an AZE observation.

\section{Experimental proposal} \label{sec:XP}

\subsection{Transition choice} \label{subsec:TransCh}

The search for a possible candidate to observe the AZE is framed by experimental
constraints.
Even if the AZE is expected to be observable on magnetic
dipolar transitions and even more effective on electric octupolar transitions,
the very long natural lifetime (of the order of one year or more) of
the excited states involved in these transitions makes them very
inappropriate to lifetime measurement. Therefore, in what follows, we
focus on demonstrating the AZE on an electric quadrupolar transition.

The first choice candidate to confirm the predictions derived for hydrogenic
atoms is the hydrogen atom itself, by transferring the atomic population to the lowest $D$-state (the
$3D$-state would play the role of the excited state
$\ket{\mathrm{e}}$), and frequently monitoring the excited state.
A major limit lies in the level scheme of hydrogen
which allows an atom in the $3D$-state to decay to
the $2P$-states by a strong dipolar transition. The lifetime of the
$3D$-state is then conditioned by its dipolar coupling to $2P$ and is
not limited by its quadrupolar coupling to $1S$. Therefore, no measurable reduction
of the lifetime due to the AZE is expected. The same problem arises with
Rydberg states, which were originally proposed as promising candidates
\cite{kofman2000acceleration} for AZE observation
due to their transitions in the microwave domain that favor
the scaling in $(1/\omega_0)^\eta$ of Eq.~(\ref{eq:JoueLaCommeKKmain})
compared to optical frequencies.

To circumvent this problem of unwanted transitions, it is then essential to identify a metastable
$D$-state, which has no other decay route to the ground state than
the quadrupolar transition. This can be found in
the alkali-earth ions like Ca$^+$ or Sr$^+$, where the lowest $D$ level is lower in energy than any $P$-level. The order of magnitude of the lifetime of these $D$-levels
ranges from 1~ms to 1~s.
The contribution to the $D$-level spontaneous emission
  rate of \emph{two-photon} decay, allowed by
  second-order perturbation theory based
  on non-resonant electric-dipole transitions, has been calculated in
  \cite{safronova10,safronova10cor} for Ca$^+$ and Sr$^+$. The
  results show that the two-photon decay channel contributes to 0.01\%
  to the lifetime of the lowest $D$-states of Ca$^+$ and Sr$^+$. As a consequence, the spontaneous emission
  from the lowest $D$-level in Ca$^+$ and Sr$^+$ can be considered to
  be due only to electric quadrupolar transition and we then focus on
  these two atomic systems in the following.

\subsection{Measurement scheme and read-out} \label{subsec:Measure}

\begin{figure}[tb]
  \centering
  \includegraphics[scale=0.35]{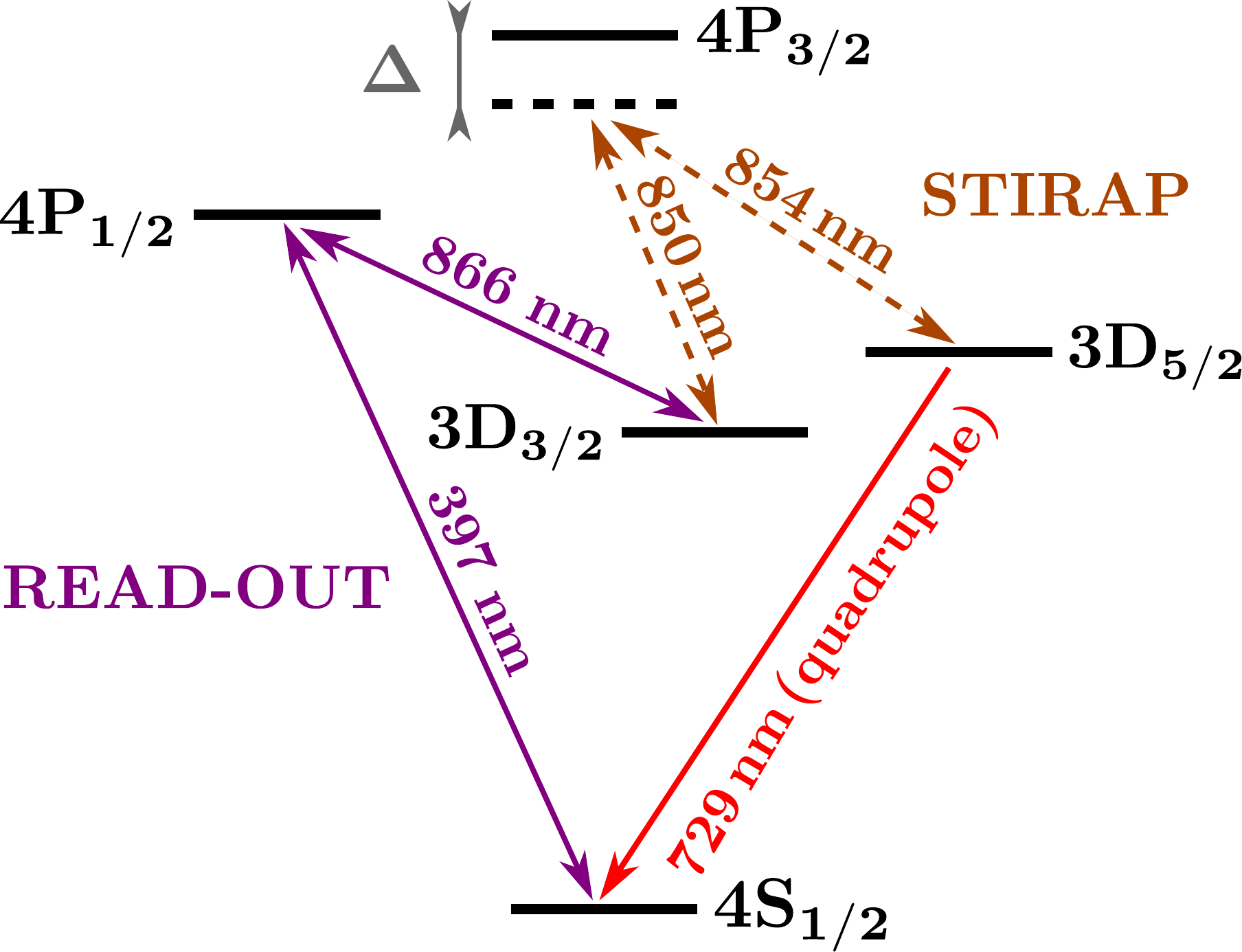}
  \caption{Dephasing measurement and read-out schemes for AZE observation in $^{40}$Ca$^+$. The
  transition used for the AZE is the electric quadrupole
  transition at $729\,\text{nm}$ (red solid arrow). The dephasing measurement
  can be performed using a STIRAP process between the
  $D_{5/2}$ and $D_{3/2}$ states \emph{via} two strong electric dipole
  transitions $D_{5/2}\to P_{3/2}$ at $854\,\text{nm}$ and $D_{3/2}\to
  P_{3/2}$ at $850\,\text{nm}$, both detuned  from
  resonance (brown dashed arrows) \cite{sorensen2006efficient}. The read-out
  consists in observation of laser induced fluorescence if the atom has decayed to the ground state \cite{dehmelt1975proposed} (purple solid arrows).}
  \label{fig:ca}
\end{figure}

Concerning the measurements of the frequently monitored excited state, 
ideal instantaneous projections on $\ket{\mathrm{e}}$ are not strictly
required. Indeed, they amount in effect to dephasing the level $\ket{\mathrm{e}}$, that is, make the
phase of state $\ket{\mathrm{e}}$ completely random
\cite{kofman2000acceleration}. Different schemes were proposed to
emulate projective measurements in
Refs.~\cite{kofman2000acceleration,kofman2001universal,ai2013quantum} and
performed in Ref.~\cite{harrington2017quantum}, for which Eq.~(\ref{eq:art_decay}) still
holds.
Here, we propose an alternative protocol in the same spirit of the ``dephasing-only measurement'' of Ref.~\cite{harrington2017quantum}.  
In this scheme, state $\ket{\mathrm{e}}$ is the metastable state $D_{5/2}$ and
the dephasing measurement is driven by the transition from $D_{5/2}$
to $D_{3/2}$, by two lasers through the strong electric dipolar
transitions to the common excited state $P_{3/2}$ (see
Fig.~\ref{fig:ca}) using a stimulated Raman adiabatic passage
(STIRAP) process \cite{kuklinski89}. If
the two-photon Raman condition is fulfilled (identical detuning for the two transitions),
the intermediate $P_{3/2}$-state is not populated and the population
is trapped in a coherent superposition of the two states
$D_{5/2}$ and $D_{3/2}$. By changing the laser power on each
transition with appropriate time profile and time delay,
the atomic population can be transferred between the two metastable
$D$-states, like demonstrated in Ref.~\cite{sorensen2006efficient}.
After one transfer and return, 
state $\ket{\mathrm{e}}$  thus acquires a phase related to the
phase of the two lasers.  By applying a random phase jump on one laser
between each completed STIRAP transfer,
the phase coherence of the excited state $\ket{\mathrm{e}}$ is washed
out, and a ``dephasing'' measurement
of the level $\ket{\mathrm{e}}$ is performed. 

To measure the effective lifetime of the $D_{5/2}$-state, the read-out
of the internal state must be based on electronic states which
do not interfere with $D_{5/2}$. 
For that purpose, the electron-shelving
scheme first proposed by Dehmelt can be used \cite{dehmelt1975proposed}.
It requires two other lasers, coupling
to the $S_{1/2} \to P_{1/2}$ and to the $D_{3/2} \to P_{1/2}$
transitions (see Fig.~\ref{fig:ca}). When shining these two lasers simultaneously, the
observation of scattered photons at the $S_{1/2} \to P_{1/2}$
transition frequency is the signature of the decay of the atom to the
ground state \cite{kreuter05}. This read-out scheme is switched on
during a short time compared to the lifetime of the $D_{5/2}$-state,
at a time when the STIRAP process has brought back the electron to $D_{5/2}$.

\subsection{Calculation for $^{40}$Ca$^+$} \label{subsec:Calcalc}

We now  try to see whether the AZE might be observable in  Ca$^+$, which is not strictly speaking
  hydrogenic, but is alkali-like in a sense that it has a single
  valence electron, and can be seen as a
  single electron orbiting around a core with a net charge
  $+2e$. Ca$^+$ is the lightest of the alkali-earth ions
having the appropriate level-scheme required for the proposed
experimental protocol (see Fig.~\ref{fig:ca}).
Therefore, we assume that it still makes sense to use
Eq.~(\ref{eq:JoueLaCommeKKmain}) (derived for hydrogen-like atoms)
and we apply it to the electric quadrupole ($\eta=3$)
transition $3D_{5/2}\to 4S_{1/2}$ to find
\begin{equation} 
\frac{\Gamma-\Gamma_0}{\Gamma_0}=A\,\frac{\nu}{\omega_0}\left(\frac{\omega_{\mathrm{X}}}{\omega_0}\right)^{2}
\end{equation}
where the numerical pre-factor $A$ cannot be computed for such an
electronic system (see Appendix~ \ref{app:Complete}
for a calculation of $A$ in the simpler case of hydrogen-like atoms). 

To be observable, the AZE must induce a lifetime reduction larger than
1\%, the best precision reached in recent
$3D_{5/2}$-lifetime measurements in Ca$^+$  \cite{kreuter05}.  We
evaluate $\omega_{\mathrm{X}}$ using
Eq.~(\ref{eq:omegacutoff}) with $n_{\mathrm{g}}=4$
and $n_{\mathrm{e}}=3$ and by replacing the atomic number $Z$ by the
effective number of charges
$Z_{\text{eff}}=2$. Using the frequency of this transition
$\omega_0=2\pi\times 411$~THz, this gives a ratio
$(\omega_{\mathrm{X}}/\omega_0)^2 \simeq 6.6\cdot10^6$. If the unknown pre-factor
$A$ is assumed  to be of the order of unity, one would
need $\nu \sim 4\,\text{MHz}$ to meet the observation requirement.

The transfer between 
the states $D_{5/2}$ and $D_{3/2}$ has been demonstrated in $^{40}$Ca$^+$ with a STIRAP process
\cite{sorensen2006efficient}, where a complete one-way transfer
duration of 5~$\mu$s was observed for 420~mW/mm$^2$ on the
850~nm  $3D_{3/2}\to 4P_{3/2}$ transition and
640~mW/mm$^2$ on the 854~nm $3D_{5/2}\to 4P_{3/2}$ transition, with both lasers
detuned by $\Delta=600$~MHz from resonance (see Fig.~\ref{fig:ca}). To
reduce the duration of the dephasing measurement
to time scale smaller than 1~$\mu$s, one can increase the laser
intensity by stronger focusing and/or larger power, but we can
also consider that a complete STIRAP transfer is not required to
achieve a dephasing of the excited state. Furthermore,  a
close inspection of Tables~\Rmnum{1} and
\Rmnum{2} in Ref.~\cite{Seke1994} suggests that the pre-factor $A$
could
be much larger than unity, making the constraint on a high measurement rate less stringent for AZE observation.

Even if the experimental requirements for AZE observation on
quadrupole transition in Ca$^+$ are more demanding than today's
best achievements, realistic arguments show that they can be met in a
dedicated experimental set-up. This experimental
challenge would benefit from theoretical insight concerning the still unknown pre-factor scaling the lifetime reduction.

\section{Conclusion} \label{sec:Ccl}

Based on well-established results for hydrogen-like atoms, we derived an analytical expression of the
decay rate modified by frequent measurements which 
allows us to highlight the main condition for an observable AZE in atomic
radiative decay in free space: all transitions except
electric-dipole transitions will exhibit an AZE under sufficiently rapid
repeated measurements.
This analytical formula also indicates how the AZE scales with the
measurement rate. We then identified a suitable level scheme in the
alkali-earth ions Ca$^+$ and Sr$^+$ for AZE observation, involving the
electric quadrupole transition between $D_{5/2}$ and $S_{1/2}$, and using a new
``dephasing'' measurement protocol based on the STIRAP technique.
Other suitable experimental schemes
might exist, and we encourage
further proposals in this sense.

\section*{Acknowledgments}

We thank Siddartha Chattopadhyay and David Wilkowski for helpful
discussions. CC acknowledges fruitful discussions with
Gonzalo Muga (UPV/EHU). EL would like to thank the Doctoral School
"Physique et Sciences de la Mati\`ere" (ED 352) for its funding.

\appendix

\section{Form of the reservoir coupling spectrum for hydrogen-like
  atoms in free space}
\label{app:reservoir}

Using the notations introduced in Sec.~\ref{subsec:ReservoirH}, the
reservoir coupling spectrum
(\ref{eq:reservoir}) is given by
\begin{multline}\label{app:eq:reservoir}
  R(\omega)=\sum_{J,M,\lambda}\hbar^{-2}\rho\left(\omega\right)\\
  \times|\bra{n_{\mathrm{e}},l_{\mathrm{e}},m_{\mathrm{e}};0}\hat{H}_I\ket{n_{\mathrm{g}},l_{\mathrm{g}},m_{\mathrm{g}};J,M,\lambda,\omega}|^2.
\end{multline}
Here, the density of states is $\rho\left(\omega\right)=1$, 
on account of the normalisation 
$\braket{J,M,\lambda,\omega\mid
  J',M',\lambda',\omega'}=\delta_{JJ'}\delta_{MM'}\delta_{\lambda\lambda'}\delta\left(\omega-\omega'\right)$ (this can be understood by dimensional considerations). 
In the non-relativistic approximation, Seke calculated in
\cite{Seke1994} the exact matrix elements
$\bra{n_{\mathrm{e}},l_{\mathrm{e}},m_{\mathrm{e}};0}\hat{H}_I\ket{n_{\mathrm{g}},l_{\mathrm{g}},m_{\mathrm{g}};J,M,\lambda,\omega}$
for hydrogen-like atoms in free space, using
the interaction Hamiltonian (in SI units)
\begin{equation} \label{eq:IntH}
\hat{H}_I=\frac{e}{m_e}\,\mathbf{\hat{A}}\left(\mathbf{\hat{x}}\right)\cdot\mathbf{\hat{p}}\;,
\end{equation}
with $e$ the elementary electric charge, $m_e$ the electron
mass, $\mathbf{\hat{x}}$ and $\mathbf{\hat{p}}$ the
position and the linear momentum operators of the
electron respectively and $\mathbf{\hat{A}}$ the vector potential operator of
the quantized EM field.
By employing these exact matrix elements (Eqs.~(17-19) in
\cite{Seke1994}) in Eq.~(\ref{app:eq:reservoir}),
one gets the following analytical form for the reservoir coupling
spectrum
\begin{multline}\label{eq:Rj}
R(\omega) = \sum_{J,M,\lambda}\hbar^{-1} (-\mathrm{i})^{2J+2\epsilon}\alpha^4 m_e c^3 \\\times
\braket{l_{\mathrm{g}},J,m_{\mathrm{g}},M|l_{\mathrm{g}},J,l_{\mathrm{e}},m_e}^2
\\\times \frac{\left(\frac{\omega}{\omega_{\mathrm{X}}}\right)^{2J+2\epsilon-1}}
{\left[
    1+\left(\frac{\omega}{\omega_{\mathrm{X}}}\right)^2\right]^{2(n_{\mathrm{g}}
    + n_{\mathrm{e}} - 1)}} \left(\sum_{r=0}^{N_J'} d_{Jr}' \left(\frac{\omega}{\omega_{\mathrm{X}}}\right)^{2r}\right)^2
\end{multline}
where $c$ the speed of light in vacuum, $\alpha$ is the fine structure
constant of electrodynamics,
$\braket{l_{\mathrm{g}},J,m_{\mathrm{g}},M|l_{\mathrm{g}},J,l_{\mathrm{e}},m_{\mathrm{e}}}$ are the Clebsch-Gordan
coefficients of the transition of interest, and $\omega_{\mathrm{X}}$ is the
non-relativistic cutoff frequency given by Eq.~(\ref{eq:omegacutoff}).
The coefficients $d_{Jr}'$ are numerical
coefficients that have been calculated for certain transitions in
\cite{Seke1994} (note that the coefficients $d_{Jr}'$ here correspond to
the coefficients $d_{00}d_{r}$ in Eq.~(18) in Ref.~\cite{Seke1994}).
The index at which the sum is
terminated is
$N_J'=n_{\mathrm{e}}+n_{\mathrm{g}}-2-(1/2)(J-l_{\mathrm{e}}-l_{\mathrm{g}}-\epsilon)$
with  $\epsilon=0$ for electric transitions and $\epsilon=1$
for magnetic transitions. Eq.~(\ref{eq:Rj}) can be recast in the form
\begin{multline}
R(\omega) = \sum_{J,M,\lambda}\hbar^{-1}
(-\mathrm{i})^{2J+2\epsilon}\alpha^4 m_{\mathrm{e}} c^3 \\\times
\braket{l_{\mathrm{g}},J,m_{\mathrm{g}},M|l_{\mathrm{g}},J,l_{\mathrm{e}},m_{\mathrm{e}}}^2
\\\times \sum_{r=0}^{N_J} d_{Jr} \frac{\left(\frac{\omega}{\omega_{\mathrm{X}}}\right)^{2J+2\epsilon-1+2r}}
{\left[
    1+\left(\frac{\omega}{\omega_{\mathrm{X}}}\right)^2\right]^{2(n_{\mathrm{g}}
    + n_{\mathrm{e}} - 1)}}
\end{multline}
where $N_J=2\left(n_{\mathrm{e}}+n_{\mathrm{g}}\right)-4-J-l_{\mathrm{e}}-l_{\mathrm{g}}-\epsilon$
and $d_{Jr}$ are combinations of the previous $d_{Jr}'$ coefficients.
Moreover, as a consequence of the conservation of the angular momentum,
the values of $J$ and $M$ must verify
\begin{equation}
\left\{
    \begin{array}{ll}
       J=|l_\mathrm{e}-l_\mathrm{g}|,|l_\mathrm{e}-l_\mathrm{g}|+1,...,|l_\mathrm{e}+l_\mathrm{g}|\\[2mm]
       M=m_\mathrm{e}-m_\mathrm{g}\equiv \overline{M}
    \end{array}
\right.
\end{equation}
which are the \emph{exact} selection rules.
Therefore, the full reservoir takes the form
\begin{multline}
R(\omega)=\sum_{J=|l_\mathrm{e}-l_\mathrm{g}|}^{|l_\mathrm{e}+l_\mathrm{g}|}\sum_{M,\lambda}
\hbar^{-1}
(-\mathrm{i})^{2J+2\epsilon}\alpha^4 m_{\mathrm{e}} c^3 \\\times
\braket{l_{\mathrm{g}},J,m_{\mathrm{g}},M|l_{\mathrm{g}},J,l_{\mathrm{e}},m_{\mathrm{e}}}^2\,\delta_{M\overline{M}}
\\\times \sum_{r=0}^{N_J} d_{Jr} \frac{\left(\frac{\omega}{\omega_{\mathrm{X}}}\right)^{2J+2\epsilon-1+2r}}
{\left[
    1+\left(\frac{\omega}{\omega_{\mathrm{X}}}\right)^2\right]^{2(n_{\mathrm{g}}
    + n_{\mathrm{e}} - 1)}}
\end{multline}
which can be recast in the expression given in the main text by
Eq.~(\ref{eq:FullReservoir}), where we introduced dimensionless coefficients
$D_{Jr}$ involving the Clebsch-Gordan coefficients and the other
constants and the sums over $M$ and $\lambda$.

\section{Derivation of the AZE scaling in the simple case of the reservoir~(\ref{eq:SimpleReservoir})}
\label{app:Simple}

Here we derive an analytical form of the integral of Eq.~(\ref{eq:art_decay}) with the
simplified form of the reservoir (\ref{eq:SimpleReservoir}).
Keeping in mind the hierarchy $\omega_0 \ll \omega_{\mathrm{X}}$, we will proceed to derive an approximate
analytical expression of the general integral
\begin{equation} \label{eq:GeneralIntegral}
I_{\eta\mu}\left(\tau\right)=\tau\int_0^{+\infty}\mathrm{d}\omega\,\frac{\omega^\eta}{\left[1+\left(\frac{\omega}{\omega_{\mathrm{X}}}\right)^2\right]^\mu}\sin\!\mathrm{c}^2\left(\left(\omega-\omega_0\right)\frac{\tau}{2}\right)
\end{equation}
in terms of which the
measurement-modified decay rate (\ref{eq:art_decay})
is straightforwardly expressed:
$\Gamma_\tau=DI_{\eta\mu}\left(\tau\right)/\omega_{\mathrm{X}}^{\eta-1}$.
Using the binomial expansion of $\omega^\eta$, we first rewrite our integral as
\begin{multline} \label{eq:FromBinomial}
I_{\eta\mu}\left(\tau\right)=\tau\sum_{k=0}^{\eta}\frac{\eta!}{k!\left(\eta-k\right)!}\omega_0^{\eta-k}\\
\times\int_0^{+\infty}\mathrm{d}\omega\,\frac{\left(\omega-\omega_0\right)^{k}}{\left[1+\left(\frac{\omega}{\omega_{\mathrm{X}}}\right)^2\right]^\mu}\sin\!\mathrm{c}^2\left(\left(\omega-\omega_0\right)\frac{\tau}{2}\right).
\end{multline}
The $k=0$ and $k=1$ terms in the sum may be treated in a specific way. Namely, we make the following
approximation of the square cardinal sine function in
$F_{\tau}(\omega)$, that is illustrated in Fig.~\ref{fig:artistic} (b):
\begin{equation} \label{eq:SincDoor}
\mathrm{sinc}^2\left(\left(\omega-\omega_0\right)\frac{1}{2\nu}\right) \simeq \left\{
    \begin{array}{ll}
       1 \text{ for } \omega_0-\pi\nu < \omega < \omega_0 +
       \pi\nu\;, \\[2mm]
       0 \text{ otherwise}.
    \end{array}
\right.
\end{equation}
This approximation is sufficient for $k=0$ and $k=1$
only, as the integrand
in (\ref{eq:FromBinomial}) decays sufficiently fast when one moves
away from $\omega_0$ so that the frequency ranges outside the door
function (\ref{eq:SincDoor}) can be ignored.
In addition to this, in the frequency range of interest here (that is, a small
range of width $\sim\nu$ centered on $\omega_0$), we can consider that
$\omega/\omega_{\mathrm{X}} \sim 0$
(which is justified by the hierarchy $\omega_0\ll\omega_{\mathrm{X}}$).
Using these approximations, we can write the low-$k$ contribution to the integral as
\begin{multline} \label{eq:AnotherWay}
\tau \sum_{k=0}^{1}\frac{\eta!}{k!\left(\eta-k\right)!}\omega_0^{\eta-k}\\
\times\int_0^{+\infty}\mathrm{d}\omega\,\frac{\left(\omega-\omega_0\right)^{k}}{\left[1+\left(\frac{\omega}{\omega_{\mathrm{X}}}\right)^2\right]^\mu}\sin\!\mathrm{c}^2\left(\left(\omega-\omega_0\right)\frac{\tau}{2}\right)\\
\simeq \tau \sum_{k=0}^{1}\frac{\eta!}{k!\left(\eta-k\right)!}\omega_0^{\eta-k}\int_{\omega_0-\pi\nu}^{\omega_0+\pi\nu}\mathrm{d}\omega\,\left(\omega-\omega_0\right)^k\\
=2\pi\,\omega_0^\eta.
\end{multline}
Now we turn to the terms for which $k\geq2$ and that will only exist
if $\eta>1$. For these terms, replacing
the square cardinal sine by a rectangle function is no longer
valid, as the growth of $\left(\omega-\omega_0\right)^k$ is not
overridden by the decrease of
$\left(\omega-\omega_0\right)^{-2}$ that comes from the square
cardinal sine, and therefore we must consider the whole frequency range.
We therefore need to find another way to approximate the integral
(\ref{eq:FromBinomial}), and we may simply replace the
square sine by its mean value $1/2$ here to get
\begin{equation} \label{eq:MeanValue}
\sin\!\mathrm{c}^2\left(\left(\omega-\omega_0\right)\frac{1}{2\nu}\right)
\simeq \frac{2\nu^2}{(\omega-\omega_0)^2}.
\end{equation}
Also note, that we should not have $\omega_0/\nu$ excessively large,
lest the square cardinal sine converges to the Dirac $\delta$
distribution, and replacing the square sine with its average value is
no longer valid.
We then compute the resulting integral, which, in the limit
$\omega_0/\omega_{\mathrm{X}}\rightarrow0$,
acceptable for the transitions that interest us, reads
\begin{multline} \label{eq:GivesBeta}
\int_0^{+\infty}\mathrm{d}\omega\,\frac{\left(\omega-\omega_0\right)^{k-2}}{\left[1+\left(\frac{\omega}{\omega_{\mathrm{X}}}\right)^2\right]^\mu}\\
\simeq\frac{1}{2}\omega_{\mathrm{X}}^{k-1}\,B\left(\frac{1-k}{2}+\mu,-\frac{1-k}{2}\right)
\end{multline}
where $B$ refers to Euler's Beta function. 
As can be checked from (\ref{eq:FromBinomial}) and
(\ref{eq:GivesBeta}), of all the contributions for $k\geq2$,
the one for which $k=\eta$ is easily the largest (this is due, again,
to the hierarchy $\omega_0\ll\omega_{\mathrm{X}}$).
As such, we can rewrite (\ref{eq:FromBinomial}) as
\begin{multline} \label{eq:Split}
I_{\eta\mu}\left(t\right)\simeq2\pi\,\omega_0^\eta+\nu\,\omega_{\mathrm{X}}^{\eta-1}\,B\left(\frac{1-\eta}{2}+\mu,-\frac{1-\eta}{2}\right)
\end{multline}
where the
second summand on the r.h.s. of (\ref{eq:Split}) will only exist for $\eta>1$. Comparison with the natural decay rate $\Gamma_0 = 2\pi R(\omega_0)
\simeq 2\pi D \omega_0^\eta/\omega_{\mathrm{X}}^{\eta-1}$ (as $\omega_0\ll\omega_{\mathrm{X}}$)
yields
\begin{multline} \label{eq:JoueLaCommeKK}
  \frac{\Gamma_\tau}{\Gamma_0}\simeq1+\frac{1}{2\pi}\frac{\nu}{\omega_0}\\
  \times\left(\frac{\omega_{\mathrm{X}}}{\omega_0}\right)^{\eta-1}B\left(\frac{1-\eta}{2}+\mu,-\frac{1-\eta}{2}\right).
\end{multline}
Note that this result had been (partially) obtained in
\cite{kofman2000acceleration}, where the authors found that for a
reservoir of the form $R(\omega)\propto \omega^\eta$ with $\eta>1$:
$\Gamma \propto \nu\,\omega^{\eta-1}_{\mathrm{X}}$, in the approximation
$\omega^\eta_0/\omega^{\eta-1}_{\mathrm{X}}\ll\nu\ll\omega_{\mathrm{X}}$ (\emph{cf.} Eq.~(20) in Ref.~\cite{kofman2000acceleration}).

\section{Derivation of the AZE scaling in the complete case of the reservoir~(\ref{eq:FullReservoir})}

\label{app:Complete}

In this section, we extend the previous result found for a reservoir
of the simple form (\ref{eq:SimpleReservoir}) to the general form (\ref{eq:FullReservoir}).
Let us first sum over $r$, and then over $J$. In the generic case, the
FGR decay rate will be, for the reservoir
coupling spectrum (\ref{eq:FullReservoir}), given by
\begin{equation} \label{eq:Gamma0J}
  \Gamma_{0J}=D_{J0}2\pi\frac{\omega_0^{\eta_J}}{\omega_{\mathrm{X}}^{\eta_J-1}}.
\end{equation}
This is true unless $D_{J0}$ vanishes. This is the case for instance of the
electric dipole transitions ($\epsilon=0$, $J=1$) between levels sharing the same principal
quantum number (see
Table~\Rmnum{1} in Ref.~\cite{Seke1994}), due to the special properties of these dipolar transitions. That $D_{J0}$ vanishes can be shown rather easily by
using the orthogonality properties of
the Gegenbauer polynomials (see Ref.~\cite{PodolskyPauling1929} for a
derivation of the momentum-space wave functions
of hydrogen in terms of these polynomials). However, we do not focus on this special case here. In the generic case
($D_{J0}\neq0$), the decay rate under frequent observations
for a specific $J$ will be
\begin{multline} \label{eq:JGammaTauJ}
  \Gamma_{\tau J}\simeq 2\pi\,D_{J0}\,\frac{\omega_0^{\eta_J}}{\omega_{\mathrm{X}}^{\eta_J-1}}+\nu\times\sum_{r=0}^{N_J}D_{Jr}\,\theta\left(\eta_J+2r-\frac{3}{2}\right)\\
  \times \,B\left(\frac{1-\eta_J-2r}{2}+\mu,-\frac{1-\eta_J-2r}{2}\right).
\end{multline}
with $\theta$ the Heaviside step function.
It thus appears that, for given $J$, all terms in the sum over $r$ in
(\ref{eq:FullReservoir}) have a
contribution to the modified decay rate that is of the same order of
magnitude. All that remains to be done is to sum over $J$.
This sum is resolved quite differently for the free decay rate on the
one hand, and the modified decay rate on the other.
Namely, for the former, we see from (\ref{eq:Gamma0J}) that the
hierarchy $\omega_0\ll\omega_{\mathrm{X}}$ ensures
that the contribution from the smallest possible $J$ is dominant.
This value is equal to
$\left|\l_{\mathrm{e}}-l_{\mathrm{g}}\right|\equiv J_{\mathrm{min}}$,
and we will write
$\eta_{\mathrm{min}}\equiv \eta_{J_{\mathrm{min}}}$. For the latter, however, we are forced to keep the double sum over $J$ and $N_J$: all (sufficiently large) powers of the frequency in the coupling contribute to the modified decay rate on the same level, with numerical prefactors as the sole difference. Namely, writing $\Gamma_0=\sum_{J=\left|l_{\mathrm{e}}-l_{\mathrm{g}}\right|}^{l_{\mathrm{e}}+l_{\mathrm{g}}}\Gamma_{0J}$ and $\Gamma_\tau=\sum_{J=\left|l_{\mathrm{e}}-l_{\mathrm{g}}\right|}^{l_{\mathrm{e}}+l_{\mathrm{g}}}\Gamma_{\tau J}$,  we have obtained
\begin{multline} \label{eq:FinalRatio}
\frac{\Gamma_\tau}{\Gamma_0}\simeq1+\frac{1}{2\pi}\frac{\nu}{\omega_0}\left(\frac{\omega_{\mathrm{X}}}{\omega_0}\right)^{\eta_{\mathrm{min}}-1}\sum_{J=\left|l_{\mathrm{e}}-l_{\mathrm{g}}\right|}^{l_{\mathrm{e}}+l_{\mathrm{g}}}\sum_{r=r_0}^{N_J}\frac{D_{Jr}}{D_{J_{\mathrm{min}}0}}\\
  \times B\left(\frac{1-\eta_J-2r}{2}+\mu,-\frac{1-\eta_J-2r}{2}\right)
\end{multline}
where we have introduced
\begin{equation} \label{eq:BeginSum}
  r_0\equiv\mathrm{max}\left\{\floor*{\frac{3}{4}-\frac{\eta_J}{2}},0\right\}.
\end{equation}
Despite the more complicated appearance of this expression, we see that the
parametric dependence of the ratio of
the decay rates is independent of the details of the
matrix elements:
the important parameter is $\eta_{\mathrm{min}}-1$. There is a
competition between $\nu/\omega_0\ll1$ and
$\omega_{\mathrm{X}}/\omega_0\gg1$
but, for $\eta_{\mathrm{min}}\geq3$, we can expect that the second
factor will dominate, especially for low values of $Z$
[see Eq.~(\ref{eq:FreqRatio})]. 
This second factor becomes all the more dominant for higher values of
$\eta_{\mathrm{min}}$, that is,
for transitions with high difference between the orbital angular
momenta of the initial and final levels.
Only for transitions where
$\left|l_{\mathrm{e}}-l_{\mathrm{g}}\right|=1$ does the second factor
$\left(\omega_{\mathrm{X}}/\omega_0\right)^0=1$ fail to play a role,
so that they verify $\Gamma_\tau\simeq\Gamma_0$.
These transitions are often called ``electric dipole transitions''
(including by us in our Sec.~\ref{sec:Intro}),
although in most cases they are accompanied by emission of photons of
angular momentum $J>1$ (as well of course as $J=1$).
Indeed, as we have recalled with Eq.~(\ref{eq:Gamma0J}), it is always
the photons with the smallest allowed value
$J_{\mathrm{min}}$ of $J$ that dominate the \emph{spontaneous} emission in an
electronic transition.

\bibliographystyle{apsrev4-1} 
\bibliography{biblio} 

\end{document}